\documentstyle[prl,multicol,aps,psfig,epsf]{revtex}
\begin{document}
\draft
\title{Magnetic Gaps related to Spin Glass Order in Fermionic Systems}
\author{R. Oppermann and B. Rosenow
\\
}\address{
Institut f\"ur Theoretische Physik,\\ Univ. W\"urzburg,
D--97074 W\"urzburg, FRG}
\date{december 97, accepted for publication}
\maketitle
\begin{abstract}
We provide evidence for spin glass related magnetic gaps 
in the fermionic density of states below the freezing temperature.
Model calculations are presented and proposed to be relevant for
explaining resistivity measurements which observe a crossover
from variable--range-- to activated behavior.
The magnetic field dependence of a hardgap
and the low temperature decay of the density of states are given.
In models with fermion transport a new metal-insulator transition 
%from hardgap insulating -- into metallic spin glass due to random 
is predicted to occur due to the spin--glass gap, anteceding 
the spin glass to quantum paramagnet transition 
at smaller spin density. Important fluctuation effects due to
finite range frustrated interactions are estimated and discussed.
\end{abstract}
\begin{multicols}{2}
\narrowtext
\tighten
Magnetic hard gaps were observed in a large number of systems such as 
In--doped CdMnTe \cite{{Terry},{Rigotty}}, amorphous GeCr--films 
\cite{GeCr}, boron-- and 
arsenic--doped silicon \cite{{Saravchik},{Dai},{Castner},{dutch}}, in 
amorphous $Si_{1-x}Mn_x$--films \cite{Ad kins}, and irradiated polymers 
\cite{{Shlimak1},{Shlimak2}} 
by manifestation of a thermal crossover from variable range hopping-- or 
Mott--behavior to activated $exp(E_H/T)$--behavior in the low T resistivity. 
Reduction of the activation energy associated with the gap under an
increased external magnetic field was held
responsible for both the generally observed feature of a negative 
magnetoresistance and the magnetic origin of the gap.
In all experiments the crossover in resistivity behavior was seen at 
temperatures close to the corresponding spin glass freezing temperature.
An existing theory involving localized magnetic polarons 
\cite{{Chicon},{Dietl}},
which explained the gradual hardening of Coulomb gaps \cite{{efshl},{Mott}}, 
was considered to be
a possible and appropriate explanation of some of the experiments 
\cite{{Terry},{Ad kins}}.
For other experiments \cite{{GeCr},{Dai},{Shlimak1},{Shlimak2}} spin glass 
order was suggested to be at the origin of the observed activated behavior. 
The random position of the implanted ions in these materials leads to a broad 
distribution of exchange integrals which favors random freezing of the
magnetic moments.\\
Fermionic spin glasses are marked by the fascinating combination of 
a complex energy landscape with phenomena well known from strongly
correlated systems like quantum phase transitions, non Fermi liquid 
behavior, phase separation, and interaction induced metal--insulator 
transitions. The paramagnet to spin glass quantum phase 
transition occuring in these systems was studied in \cite{SRO,Georges,brro}, 
while strong correlation effects deep
inside the spin glass phase were discarded. 
Recently the properties of highly correlated metals and interaction--induced
insulators have attracted much interest \cite{GeKr}.
%Comparable phenomena in disordered systems are not equally well 
%investigated despite the fact that the absence of a notorious 
%coupling between low and high energy modes simplifies  
%calculations and renders an analytical treatment feasible.
\\
The goal of this Letter is to report theoretical results, obtained by 
replica--methods and TAP--equations, for comparable phenomena in several 
different fermionic spin glasses, which develop either a hard gap or a 
pseudogap in the fermionic density of states at zero temperature, 
depending in size and shape on the spin glass order parameter(s).  
In any case the manifestation of spin glass order in this fermionic property
represents a surprisingly large effect.\\
The magnetic field induced reduction of the gap, 
implying a negative magnetoresistance, and its stability against fermion 
hopping transport are demonstrated. 
Since the system stays insulating due to spin 
glass order it may be characterized as a SG (spin glass) based  insulator.
The existence of a SG based  metal insulator transition is conjectured
from the combination of our new calculations with results obtained
in the vicinity of the quantum paramagnet to spin glass transition in
metallic systems \cite{OpBi}.
The low energy behavior of the Green function in the SG based insulating phase 
is derived and an estimation for the location of the abovementioned 
transition is given.  We discuss implications of
our results for the interpretation of the experiments 
\cite{{Terry},{Rigotty},{GeCr},{Saravchik},{Dai},{Castner},{dutch},{Ad
kins},{Shlimak1},{Shlimak2}}.\\
%*************model definition**************************
The fermionic Ising spin glass $ISG_f$  is defined by the Hamiltonian
${\cal{H}}=-\frac{1}{2}\sum J_{ij}\sigma_i\sigma_j-\mu\sum
n_i$ with spins $\sigma =n_{\uparrow}-n_{\downarrow}$, the
fermion number operator $n=n_{\uparrow}+n_{\downarrow}$,
chemical potential $\mu$
, and Gaussian--distributed 
$J_{ij}$ with variance $J^2$.
%**************************************************************
It is the simplest model that 
takes charge--spin couplings and corresponding fluctuations into account 
and, by virtue of its exact solvability in infinite space dimensions and
for $T\rightarrow 0$ provides solid 
theoretical background to be compared with related experimental situations.
Due to the possibility of double and zero occupancy of sites the freezing 
temperature is slightly lower than the one of its 
standard counterpart (SK--model) with one spin at each site.
Thermally activated hopping of the localized fermions is effectively 
allowed for by 
working in the grand canonical ensemble with statistically fluctuating 
particle numbers at each site.
While spin correlation functions of the model are static, 
correlators with an odd number of equal time fermion operators display
the rich quantum--dynamical structure of an interacting fermion system. 
The fermionic path integral technique provides the tool for the calculation
of quantities such as density of states, (dc or ac) conductivities et cetera 
for all kinds of spin glass models with charge degrees of freedom. 
We consider important the question, whether such a purely magnetic 
model, whose Hamiltonian consists just of a frustrated 
spin interaction, can have specific and strong effects on charge 
correlations and fluctuations, and whether a continuous distribution of 
exchange energies can cause a rigid gap of finite width.\\
The density of states $\rho_{\sigma}(\epsilon)=- \frac{1}{\pi} Im 
G_{\sigma}^R(\epsilon)$, 
calculated by means of the replica--method from the fermion Green's function 
of the $ISG_f$ , reads 
%>>>>>>>>>>>>>>>>>>>>>>>>>>>>>>>>>>>>>>>>>>>>>>>>>>>>>>>>>>>>>>>>>>>>>>>>>>>>
%                       EQUATION 1
%============================================================================
\begin{eqnarray}
& &\rho_{\sigma}(\epsilon)
=\frac{ch(\beta\mu)+ch(\beta(\epsilon+\mu))}{\sqrt{2\pi 
T\chi}\hspace{.1cm}J}
\label{one}\\
\nonumber & &\hspace{-.5cm}\int_{-\infty}^{\infty}\frac{dz}{\sqrt{2\pi}}
\frac{exp\left[-\frac{z^2}{2}-\beta
(\epsilon+\mu+\sigma\tilde{H}(z))^2)/(2 J^2 \chi)\right]}
{ch(\beta\mu)+exp(\frac{1}{2}\beta J^2\chi)ch((\beta\tilde{H}(z))}\ ,
\end{eqnarray} 
%<<<<<<<<<<<<<<<<<<<<<<<<<<<<<<<<<<<<<<<<<<<<<<<<<<<<<<<<<<<<<<<<<<<<<<<<<<<<
where $\tilde{H}(z)=H+J\sqrt{q}z$,$H$, $q$, $\chi$
denote effective field, external magnetic field, spin glass 
order parameter, and linear susceptibility. 
Effects from replica symmetry breaking (RSB), involving 
$\infty$--many order parameters $q_i$ and variables $z_i$, 
and from finite--range interactions $J_{ij}$
are not yet included here, but are considered below.
%While $q, \chi$, and $\nu(\mu)$ can 
%be obtained by extremizing the free energy, ie without explicit knowledge 
%of the underlying density of states (DoS), this fermionic DoS     
%is of course related to the magnetic properties of the system and its 
%analysis requires the solution of several selfconsistency
%equations for spin-- and for charge correlations. 
%We obtained these solutions for low temperatures and for the 
%entire range of magnetic fields and single fermion excitation energies.
The existence of a hardgap of size $2E_g(H)$ is nevertheless an unexpected 
and surprisingly large effect of spin glass order; it is best appreciated in our
result for the density of states result at $T=0$, which is given by
%>>>>>>>>>>>>>>>>>>>>>>>>>>>>>>>>>>>>>>>>>>>>>>>>>>>>>>>>>>>>>>>>>>>>>>>>>>>>
%                       EQUATION 2
%============================================================================
\begin{equation}
\rho_{\sigma}(E)|_{_{T=0}} = \frac{^{\Theta(|E| - 
E_g(H))}}{_{\sqrt{2\pi}J}}e^{^{-\frac{1}{2 J^2}\left[E
-E_g(H)sgn(E)+\sigma H\right]^2}},
\label{2}
\end{equation}
%<<<<<<<<<<<<<<<<<<<<<<<<<<<<<<<<<<<<<<<<<<<<<<<<<<<<<<<<<<<<<<<<<<<<<<<<<<<<
with $E:=\epsilon+\mu$ and the field--dependent gap energy $E_g(H) =J 
\sqrt{2/\pi} exp(-H^2/(2 J^2))$.  
Eq.(\ref{2}) is valid within the regime
$|\mu|<E_g/2$, which corresponds to a half filled system at $T=0$
with filling factor $\nu=1+(1-q-T\chi)tanh(\beta\mu)$.
In terms of $E_g(H)$ the following low temperature expansions 
(for linear susceptibility and order parameter), 
required for the exact evaluation of the DoS formula (\ref{one}) at low T, 
were obtained as
$q= 1 - E_g(H)T + O(T^2),\chi= E_g(H) + \frac{1}{2} E^2_g(H)(1-H^2)T + O(T^2)$.
The last equation allows one to express the zero field gapwidth as $E_g=
a(\mu) \chi T_c^2$ completely in terms of
experimentally accessible quantities, where $a$ increases with the chemical 
potential from $2.1833$ at $\mu=0$ to $2.4075$ at $\mu=J/\sqrt{2\pi}$. 
These results show that i) the density of states is zero at $T=0$ in the
finite interval given by $|\epsilon+\mu|<E_g(H)$ and ii) the gapwidth 
%****************** changes below ***********************************
in a magnetic field is independent of spin orientation (in contrast to the
local--limit Hubbard gap in an external field) and shrinks as 
the field is increased. 
%********************* changes above ********************************
This will be the source of the negative 
magnetoresistance in the extended Ising spin glass model with charge transport 
as discussed below. The physical consequences of these spin glass related 
properties are hence in agreement with the experimental observations of 
crossover behavior in the low T resistivities mentioned above. 
For single fermion energies $E\equiv \epsilon+\mu$ smaller than the 
gap energy $E_g(H)$ and $|\mu| < \frac{1}{2}E_g(H)$
the density of states decays to zero exponentially as given by
%>>>>>>>>>>>>>>>>>>>>>>>>>>>>>>>>>>>>>>>>>>>>>>>>>>>>>>>>>>>>>>>>>>>>>>>>>>>
%                       EQUATION 3
%===========================================================================
\begin{eqnarray}
& &\hspace{-.5cm}\rho_{\sigma}(E)= 
\frac{1}{_{2\hspace{.1cm}J\hspace{.1cm}
cos(\frac{\pi}{2}\frac{E}{E_g(H)})}}\hspace{.1cm} 
e^{-\frac{1}{4}\frac{E^2_g(H)-E^2}{J^2}(1-\frac{H^2}{J^2})
-\frac{1}{2}\frac{H^2}{J^2}}\nonumber\\
& &(\beta E_g(H))^{-\frac{1}{2}}\hspace{.1cm}
\left[ch(\beta\mu)+ch(\beta E)\right]\hspace{.1cm}
e^{-\frac{1}{2}\beta\frac{E^2_g(H)+E^2}{E_g(H)}}.
\end{eqnarray}
%<<<<<<<<<<<<<<<<<<<<<<<<<<<<<<<<<<<<<<<<<<<<<<<<<<<<<<<<<<<<<<<<<<<<<<<<<<<
\begin{figure}
\vspace{-2cm}\hspace{.7cm}
\psfig{file=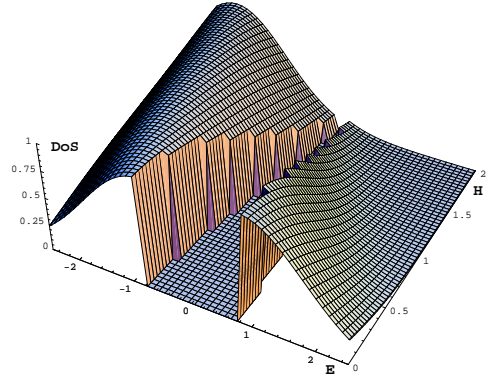,width=6.5cm,angle=0}
\vspace{-1.5cm}
\caption{Single particle density of states (DoS) for the fermionic Ising 
spin glass
($ISG_f$) versus energy and magnetic field in units of J. 
%The hardgap centered around zero
%energy is due to random magnetic correlations and manifests itself in an
%activated behavior of the low temperature hopping conductivity of 
%disordered
%magnetic materials. The reduction of the gapwidth by an external magnetic 
%field
%explains the observed negative magnetoresistance.
}
\end{figure}  
%#######################################################################
The infinite range insulating spin glass model allows to evaluate 
effects of replica symmetry breaking (RSB). 
Let us discuss the ($T=0,H=0$) result for the fermion Green's function 
%>>>>>>>>>>>>>>>>>>>>>>>>>>>>>>>>>>>>>>>>>>>>>>>>>>>>>>>>>>>>>>>>>>>>>>>>
%                        EQUATION 4
%=======================================================================
\begin{equation}
G^R(E)=\sum_{\lambda=\pm1}\int_{0}^\infty d\tilde{H}\;
P(\tilde{H})\;
\frac{1}{_{E+i0-\lambda
\left[\tilde{H}+\bar{\chi}\right]}}\ \ .
%+
%\frac{1}{_{E+i0+\tilde{H}+\bar{\chi}}}]
\label{4}
\end{equation}
%<<<<<<<<<<<<<<<<<<<<<<<<<<<<<<<<<<<<<<<<<<<<<<<<<<<<<<<<<<<<<<<<<<<<<<<
The weight $P(\tilde{H})$ depends on the set of Parisi
order parameters and  
simplifies to $\frac{1}{\sqrt{2\pi q}}e^{-\frac{\tilde{H}^2}{2 q}}$
under replica symmetry reproducing thus Eq.(\ref{2}) together with 
$\overline{\chi}\equiv \chi$. 
Hence the one particle excitations reveal a gap if the nonequilibrium 
susceptibility $\bar{\chi}$, which enters as a self energy and represents 
the effect of the Onsager 
reaction field, is finite. The influence of the randomly distributed local 
fields on the other hand is contained in the weight $P(\tilde{H})$. For 
the infinite--range model, the RSB--solution 
shows that $\bar{\chi}$ vanishes like T for $T\rightarrow0$. Our explicit
calculations show that in this case the 
hardgap turns into a pseudogap (details will be published elsewhere). 
Finite range interactions however lead to corrections that 
render $\bar{\chi}$ finite \cite{BiYo}, which indicates that 
fluctuations will tend to reinstall the gap. 
Whereas a controlled theory for self energy corrections to Eq.(\ref{4})
does not yet exist, an estimation of these small fluctuation 
contributions to the density of states is given below.\\
%********************** Changes below******************
%We find that the hardgap persists under one--step replica symmetry 
%breaking (RSB). Its width is seen to be related to the 
%nonequilibrium susceptibility $\bar{\chi}$ 
%obtained from the solution of the TAP--equations as. 
%In this context fluctuation effects favoring the gap in real 
%systems with finite--range interactions become very important as 
%discussed below. 
%************************* changes above **********************
%
The gap also turns out to be robust against an additional fermion hopping 
term until the bandwidth exceeds a critical value. 
The gap and related properties of this extended itinerant spin glass 
model, whose magnetic phase diagram and --transitions have 
previously been analyzed \cite{{SRO},{brro}}, are also discussed below.\\
Coulomb interaction effects are of diverse nature: the long range part
will still tend to depress the (remaining) DoS near the Fermi level and
thus stay particularly relevant when either
spin glass order is weak or absent, or (in the case of fully developed spin 
glass order) when the Fermi energy lies close to the gap edges.
The Hubbard coupling U leads to a shift of the chemical potential
($\mu\rightarrow \mu-\frac{U}{2}\nu$), and also to a shift 
$H\rightarrow H^{\prime}\equiv H+\frac{U}{2} m(H)$
of the applied magnetic field.\\
Corrections associated with a finite interaction range are estimated
by considering TAP--equations for the $ISG_f$. 
Similarly to the spin case an Onsager reaction
term $-\sum_{j=1}^z J_{ij}^2 \sigma_i \chi_{jj} m_i$ (coordination number z)
must be subtracted from the local field on the level of frozen in
magnetizations, but to account for spin fluctuations a self 
energy  $\frac{1}{2}\sum_{j=1}^z J_{ij}^2 \sigma_i \chi_{jj} \sigma_i$ must be 
added again. 
%---------------------
Thus the polarization $\sigma_i J_{ij} \chi_{jj}$ of surrounding spins
lowers the energy of magnetic sites by $E_g= [  \sum_{j=1}^z \chi_{jj}
(J_{ij})^2]_{av}$.
As this gap energy originates in the formation of polarized spin
clusters it
differs from the Hubbard gap width U caused by onsite repulsion.
%---------------------
%In this way the energy of magnetic states is lowered by
%$E_g= [  \sum_{j=1}^z \chi_{jj} (J_{ij})^2]_{av}$.
Now we argue that for a finite interaction range the local susceptibility 
behaves like $\overline{\chi} = const. +O(T)$ instead of the 
$\tilde{\chi} \sim T$ found by solving the TAP--equations for the
infinite--range SK--model \cite{TAP}.  The latter is valid for a system 
stuck in a fixed
free energy valley with infinitely high barriers and is intimately related 
to a pseudogap $P(h_{loc})\sim h$ in the distribution of local fields, which
is absent in finite dimensional systems \cite{BiYo}.
Thus fluctuation effects in systems 
with finite--range 
frustrated interactions must be expected (despite the not yet 
available finite--range spin glass theory) to favor the hardgap as predicted
by the replica symmetric solution eq.(\ref{2}) of the infinite range model.
Hence we conjecture
$\rho(\epsilon)\sim [\Theta( \epsilon - \chi  \sum_{j=1}^z (J_{ij})^2)]_{av}
\sim \epsilon^{\frac{z}{2}}$ and therefore along the lines of \cite{efshl}
$\sigma_{DC} \sim \exp(-B/T^{\frac{z+2}{z+8}})$ for bonds with a Gaussian
distribution. A sufficiently large 
(magnetic) coordination number z or a long--ranged RKKY--interaction leads
to an activated behavior of the conductivity, otherwise
nonuniversal temperature exponents may occur. A bimodal
$\pm J$ -- distribution on the other hand is expected to result in a true 
hardgap.\\
Multi--valley correlations of statistical DoS - fluctuations as given by 
$C_{\rho}^{(k)}\equiv
\overline{\rho_{a_1}(E_1)\rho_{a_2}(E_2)...\rho_{a_k}(E_k)}$, 
where the $a_i$ denote distinct replicas, do not factorize within the 
spin glass phase. For $|E_k|=E<E_g(H)$ we obtain
$C_{\rho}^{(k)}(E)=
(\rho(E))^k\frac{B(\frac{k}{2} d_-^E,\frac{k}{2} d_+^E)}
{(B(\frac{1}{2}d_-^E,\frac{1}{2}d_+^E))^k}$ with 
$d_{\pm}^E\equiv \frac{E_g(H)\pm E}{E_g(H)}$ and Beta--function B.
These multi--valley correlations also vanish 
exponentially within the gap as $T\rightarrow 0$ and, 
weaker than the averaged DoS though, 
they increase as either gap edge is approached at fixed low T.\\
%!!!!!!!!!!!!!!!!!!!exclude below!!!!!!!!!!!!!!!!!!!!!!!!!!
%For values of the chemical potential larger than half the gap energy
%we derived, in addition to the trivial paramagnetic solution, a solution 
%characterized by a continuously decreasing order parameter q and a 
%continuously increasing fermion filling $\nu$. 
A comparison of the free energy of ordered and disordered phase
predicts a first order transition to the paramagnetic phase at $\mu=0.9$,
but when analyzing the stability of the magnetic solution according to the 
scheme of de Almeida and Thouless \cite{BiYo} we found two negative 
eigenvalues of the Hessian matrix instead of the expected one negative value 
indicating just the instability towards RSB.
As described in \cite{BiYo} for spin glass problems, one has to choose
the {\it stable} solution with the lowest free energy, hence the system 
undergoes a first order transition from half filling to 
the completely filled paramagnetic state at $\mu = \frac{1}{2} E_g$.\\
The electronic kinetic energy is properly embedded in an itinerant 
$ISG_f$--model by adding the random hopping Hamiltonian
$H_t= 
\sum_{i,j;\sigma} t_{ij} c_{i,\sigma}^{\dagger} c_{j,\sigma}.$
The real--symmetric $t_{ij}$ are Gaussian--distributed and chosen to 
describe a semi--circular band of width $E_0$.
The 4--fermion term which results from 
$t_{ij}$--averages is decoupled by means 
of a quaternionic matrix field $\underline{\underline{R}}$ in a way familiar
from Anderson localization theory. The saddle point value
$r_{\sigma}(\epsilon_l)$ of this field obeys the selfconsistency equation
$r_{\sigma}(\epsilon_l)=
\frac{i E_0^2}{4} G_{\sigma}(\epsilon_l) = 
-\frac{i E_0^2}{4n} \sum_a\int_0^\beta d\tau e^{i\epsilon_l \tau}
<T_{\tau}c_{a,\sigma}(\tau)c_{a,\sigma}^{\dagger}(0)>$, which becomes exact 
for infinite space dimensions. 
The expectation value employs the quantum--dynamical action
${\cal A}= 
\int_0^{\beta}\int_0^{\beta} 
d\tau d\tau^{\prime}(\sum_{a,\sigma}
c_{a,\sigma}^{\dagger}(\tau)[\delta(\tau-
\tau^{\prime})(\partial_{\tau}+\mu) + i r_{\sigma}(\tau,\tau^{\prime})]
c_{a,\sigma}(\tau) +\sum_{a,b}
S^a(\tau)S^b(\tau^{\prime})Q^{ba}(\tau^{\prime},\tau))$
containing the disorder averaged spin interaction. 
In contrast to the quaternionic charge fields 
$\underline{\underline{R}}$,
$\underline{\underline{Q}}(\tau^{\prime},\tau)$ denotes the quantum spin 
glass order parameter field; 
quantum dynamics necessitate the solution of dynamical
selfconsistency equations for the spin autocorrelation function 
$<Q^{aa}(r,\tau^{\prime},\tau)>$ at saddle point level.
Apart from the different interaction term in the Lagrangian, 
the selfconsistency equation for $r_{l,\sigma}$ resembles the LISA
equation \cite{GeKr} for a model with nonrandom nearest neighbor 
hopping on a Bethe lattice with infinite coordination number, commonly used 
to study strong correlations in the Hubbard model \cite{GeKr}. 
Quantum dynamics of the $\underline{\underline{Q}}$--fields 
can be important close to the $T=0$ paramagnet - SG-- 
transition; yet the influence of the 
fermion kinetic energy on the DoS is described sufficiently well by a 
static, replica symmetric approximation 
for the spin autocorrelation function. For $H=0$ 
and half filling the Green function follows from
%>>>>>>>>>>>>>>>>>>>>>>>>>>>>>>>>>>>>>>>>>>>>>>>>>>>>>>>>>>>>>>>>>>>>
%                      EQUATION 5
%<<<<<<<<<<<<<<<<<<<<<<<<<<<<<<<<<<<<<<<<<<<<<<<<<<<<<<<<<<<<<<<<<<<<
\begin{equation}
r_l= \frac{1}{4}E_0^2 <(\epsilon_l+
r_l)/((\epsilon_l+r_l)^2+\tilde{H}^2)>_{\Phi}
\label{seven}
\end{equation}
%====================================================================
with
%\begin{equation}
$<f(\tilde{H})>_{\Phi} =\int_z^G f(\tilde{H})
\int_y^G 
\frac{e^{\Phi(\tilde{H}(z,y))}}{\int_{\tilde{y}}^G e^{\Phi(\tilde{H}(z,
\tilde{y}))}}$. 
The regularized determinant $\phi(\tilde{H})= 
Tr \log[ 1 + \frac{\tilde{H}^2}{(\epsilon_l+r_l)^2}]$ 
results from integrating out the decoupled fermion fields, and 
$\tilde{H}(z,y)=J(\sqrt{q} z + \sqrt{T \chi} y)$.
In order to proof the stability of the hard gap against fermionic hopping 
we present
a solution of the above equation valid in the low energy regime. The leading
contribution in a double expansion in $E_0$  and frequency $\epsilon_l$ is 
$r_l= \epsilon_l \frac{E_0^2}{4}<\frac{
1}{\tilde{H}^2}>_{\Phi{ISG_f}}$
giving just back  the $ISG_f$ - result.
Now we note that the ansatz $r_l=\alpha \ \epsilon_l$ i)satisfies  
eq.(\ref{seven})
to all orders in $E_0^2$ and to $O(\epsilon_l)$ for $\alpha=\frac{E_0^2 <J^2/
\tilde{H}^2>_{\Phi_{\alpha}}}{4 J^2 - E_0^2 <J^2/\tilde{H}^2>_{\Phi_{
\alpha}}}$ and ii)  corresponds to a  real $G^R(\epsilon)
=-\frac{4\alpha \epsilon}{E_0^2}$ and hence to a  vanishing DoS. Solving a 
set of coupled self consistency equations
one obtains $q_{\alpha}=1 + O(T)$, $\chi_{\alpha}= \frac{\sqrt{2/\pi}}{1+
\alpha} + O(T)$, and finally $<\frac{1}{\tilde{H}^2}>_{\Phi_{\alpha}}= 
<\frac{1}{\tilde{H}^2}>_{\Phi_{ISG_f}}=0.568659$. The last result 
implies that for $E_{0c}=2.65219 J$ the low energy approximation breaks down
since $\alpha$ diverges.
On the other hand it is known that for $E_0=32 J/(3\pi)$ the 
system undergoes a paramagnet to spin
glass quantum phase transition and has a semielliptic DoS at zero temperature
\cite{OpBi},
hence we conclude that the divergence of $\alpha$ at $E_{0c}$ indicates the
neighborhood of a metal insulator transition exclusively caused by random 
interactions.
A mobility edge induced by either randomness in fermion hopping or
in the interaction can interfere with the SG--based MIT. 
A transition between gapped and non--gapped insulating
behavior appears possible too and requires further analysis. 
Several different lower critical dimensions such as the one for Ising spin 
glass order at $T_c>0$ and the one for Anderson localization will also 
play a role in this context. Anderson localization of states 
neighboring a gap had been studied in Ref.(\cite{Polishchuk})\\
Using these results for the interpretation of experiments one should
distinguish two classes of systems. In $GeCr$ \cite{GeCr}, irradiated
polymers \cite{{Shlimak1},{Shlimak2}}, $SiB$ \cite{Dai}, ion implanted 
\cite{dutch} and Czochralski grown \cite{Castner} $SiAs$ localized impurity 
spins are the main source of magnetic behavior. The interaction between them
is due to direct wave function overlap, hence the 
spatial randomness in their positions translates itself into a broad 
distribution of exchange energies favoring spin glass freezing in the 
ground state. We believe that this experimental situation is well
described by the $ISG_f$  and that the observed activated
conductivity is explained by the spin freezing induced hard gap in the DoS.
However, in Indium--doped $CdMnTe$ \cite{Terry} and annealed $SiMn$ 
\cite{Ad kins}
films the carrier spins interact with Manganese spins, may form magnetic
polarons and thus create a finite activation energy for hopping transport.
In \cite{Terry,Ad kins} this mechanism is held  
responsible for the observed low T behavior of the conductivity. Since
$CdMnTe$ is a prototype insulating spin glass 
a SG--origin of the activation energy can only be ruled 
out by checking whether the thermal crossover of the
conductivity (in In--doped material) appears already above the freezing 
temperature.
If this were not the case an explanation of these experiments could 
be found in a spin glass of randomly interacting magnetic polarons.
Other mechanisms discussed in the literature as cause for the observed
activated conduction are a hardening of the Coulomb gap by formation of 
electronic polarons \cite{Chicon} and nearest neighbor hopping. The former 
can be ruled out because of the observed negative magnetoresistance, the 
latter because of Efros--Sklovskii variable--range hopping above the 
crossover temperature. The possibility of a dimensional crossover at 
low T needs to be considered, but it would lead to a 
$(T_0/T)^{1/3}$ behavior and not to the appearance of a hardgap.  \\
We acknowledge support by the SFB410 of the DFG. We thank
G. Landwehr and C. Rigaux for valuable hints, and a
referee for bringing ref.(22) to our attention.
%\begin{thebibliography}{99}

\end{multicols}
\end{document}